# Prime Event Languages: An Information-Theoretic Framework for Arithmetic Event Sequences

**Jinhua Liao**

Independent Researcher, McLean, Virginia, USA

Email: jinhualiao@yahoo.com

## Abstract

This work introduces the concept of a Prime Event Language, a representational framework in which arithmetic phenomena are transformed into symbolic event sequences and analyzed using information-theoretic methods. Rather than studying prime-derived phenomena solely as numerical objects, the framework treats them as event-generating processes whose statistical structure can be investigated through entropy, Markov modeling, cross entropy, mutual information, and information-horizon analysis.

Using all primes up to N = 5 x 10^9 (234,954,223 primes), we construct event languages based on twin-prime occurrences and record-gap events. For the Twin Prime Event Language, first-order Markov modeling reduces test-set cross entropy from 0.325350 bits to 0.319949 bits, corresponding to an information gain of approximately 0.0054 bits. This gain survives out-of-sample validation and therefore reflects reproducible statistical structure rather than overfitting.

Mutual-information analysis independently confirms the Markov results and shows that measurable dependence is concentrated almost entirely at lag 1. The mutual information decreases from approximately 5.96 x 10^-3 bits at lag 1 to approximately 5.07 x 10^-7 bits at lag 2, a reduction of more than four orders of magnitude. Beyond lag 2, residual information fluctuates near the statistical noise floor.

Beyond the specific empirical findings, the principal contribution of this work is representational. The Prime Event Language framework demonstrates how arithmetic phenomena may be studied not only as numerical sequences but also as symbolic and informational objects. The results suggest that alternative representations can expose measurable organizational structure that remains less visible in conventional numerical descriptions.

## 1. Introduction

Prime numbers occupy a central position in mathematics and have been studied extensively through analytic, probabilistic, and computational approaches [1,2,3,4]. Classical investigations have focused on prime density, prime gaps, asymptotic distributions, and conjectures concerning special prime configurations such as twin primes [1,2]. Despite substantial progress, many aspects of prime distributions continue to exhibit behavior that is difficult to distinguish from randomness [1,3].

The present work adopts a different perspective. Rather than analyzing prime-derived phenomena solely as numerical sequences, we investigate whether certain arithmetic events can be represented as symbolic languages and studied using tools from information theory, stochastic processes, and symbolic dynamics [5,6,7,9]. Examples of such events include the occurrence of twin-prime gaps and the appearance of record prime gaps. When represented as binary symbolic sequences, these arithmetic phenomena become amenable to analysis using entropy, cross entropy, Markovity, mutual information, and information propagation.

The motivation for this approach arises from a broader scientific observation: many advances in science occur not because new data become available, but because familiar phenomena are represented in new ways. Alternative representations often reveal organizational principles that remain hidden in conventional descriptions [8,9,10]. The present work is motivated by the question: what representations make hidden structure visible?

The central question investigated in this work is straightforward: can prime-derived event languages reveal measurable information structure? The objective is representational rather than conjectural. We do not attempt to prove new results about the distribution of primes. Instead, we ask whether an event-language representation exposes statistical organization that becomes measurable through information-theoretic analysis.

To address this question, we introduce the concept of a Prime Event Language. Two event languages are considered in the present study. The first is the Twin Prime Event Language, in which an event occurs whenever a prime gap of two is observed. The second is the Record Gap Event Language, in which an event occurs whenever a newly observed prime gap exceeds all previously observed gaps.

Using all primes up to N = 5 x 10^9 (234,954,223 primes), we investigate whether measurable dependence exists within these event languages. Finite-order Markov models quantify memory effects [7]. Train/test validation distinguishes reproducible structure from overfitting. Mutual-information analysis measures information propagation across lag distances [5,6]. Finally, information-horizon analysis characterizes the effective range over which predictive information persists.

## 1.1 Contributions

1. Introduction of the Prime Event Language framework, which represents arithmetic phenomena as symbolic event sequences suitable for information-theoretic analysis.
2. Demonstration that the Twin Prime Event Language exhibits weak but statistically reproducible first-order dependence.
3. Verification that the observed dependence survives out-of-sample train/test validation.
4. Identification of a highly localized information structure characterized by an effective information horizon of approximately one event.
5. Illustration of how alternative representations can reveal information-theoretic organization within arithmetic phenomena.

## 1.2 Related Work

The present study lies at the intersection of prime-number statistics, information theory, symbolic dynamics, and stochastic-process analysis. While each of these fields has a substantial literature, their combined application to prime-derived event languages appears largely unexplored.

### Prime Numbers and Prime-Gap Statistics

The statistical properties of prime numbers have been studied extensively for more than a century. Classical work by Hardy and Littlewood established influential conjectures regarding prime constellations and twin primes [2]. Cramer later introduced a probabilistic model of prime gaps that remains a foundational reference in the study of apparent randomness within prime distributions [1]. Modern developments, including bounded-gap results and related advances, have further demonstrated that prime gaps exhibit rich statistical structure despite the deterministic nature of the underlying arithmetic process [4].

Traditional investigations primarily focus on numerical properties such as prime density, gap distributions, asymptotic laws, and the occurrence of special configurations [1,2,3,4]. The objective is typically explanatory or predictive, seeking theoretical mechanisms responsible for observed arithmetic phenomena. In contrast, the present work does not attempt to derive new asymptotic results or establish new number-theoretic theorems. Instead, it investigates whether prime-derived phenomena exhibit measurable information structure when represented as symbolic event sequences.

### Information Theory

The analytical framework employed in this study originates from information theory. Shannon's foundational work introduced entropy as a quantitative measure of uncertainty and established the theoretical basis for information transmission and representation [5]. Subsequent developments by Cover and Thomas provided a comprehensive framework for entropy, mutual information, cross entropy, and related measures of statistical dependence [6].

Information-theoretic methods are now widely applied across communication systems, statistical learning, natural language processing, and complex systems analysis. In the present work, these methods are employed not to study communication signals but to investigate arithmetic event sequences generated from prime-gap data. Entropy, cross entropy, and mutual information provide quantitative measures of uncertainty and dependence that are independent of any specific number-theoretic model [5,6].

### Symbolic Dynamics and Sequence Representation

The event-language representation used in this work is closely related to symbolic dynamics, in which complex systems are represented as sequences of discrete symbols and analyzed through their statistical properties [9]. Symbolic representations have proven valuable in studying dynamical systems because they transform continuous or numerical processes into discrete sequences that are amenable to information-theoretic analysis.

Related ideas also appear in the theory of automatic sequences and formal language representations of mathematical objects [8]. Such approaches emphasize that alternative representations may reveal organizational properties that remain less apparent in the original formulation. The Prime Event Language framework follows a similar philosophy by transforming arithmetic phenomena into symbolic event sequences.

### Algorithmic Information and Complexity

The present work is also conceptually related to algorithmic information theory, which investigates the complexity and informational content of mathematical objects [10]. Although the current study does not directly employ Kolmogorov complexity, both approaches share a common objective: understanding mathematical structures through informational rather than purely numerical descriptions.

From this perspective, arithmetic systems may be viewed not only as collections of numbers but also as generators of information-bearing sequences. The Prime Event Language framework provides one possible mechanism for exploring this viewpoint.

### Distinction from Prior Work

The present work differs from previous studies in both objective and methodology. Traditional investigations of prime numbers typically seek asymptotic laws, probabilistic models, or theoretical explanations for observed distributions [1,2,3,4]. By contrast, the Prime Event Language framework investigates whether arithmetic phenomena possess measurable information structure when represented as symbolic event sequences.

To the best of the author's knowledge, the specific combination of event-language representation, finite-order Markov modeling, out-of-sample validation, mutual-information analysis, and information-horizon measurement has not previously been applied to prime-derived arithmetic events at the scale considered here. The novelty of the present work therefore lies not in the introduction of new information-theoretic methods, but in the application of established information-theoretic tools to a new representational framework for arithmetic phenomena.

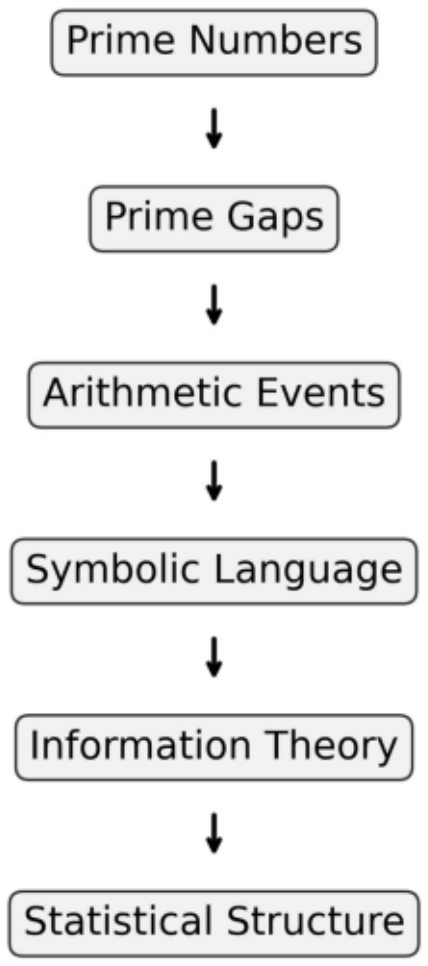

*Figure 0. Conceptual framework of the Prime Event Language approach. Prime numbers are transformed into prime gaps, arithmetic events, symbolic languages, and finally information-theoretic measurements.*

# 2. Prime Event Language Framework

## 2.1 From Arithmetic Sequences to Event Languages

Traditional studies of prime distributions focus on numerical quantities such as prime values, prime gaps, densities, and asymptotic behavior [1,2,3]. In contrast, the present work adopts an event-based representation. Let g(n) = p(n+1) - p(n) denote the gap between consecutive primes. An event language is generated by defining a rule that maps each observed gap to a symbolic value. In the present study, binary languages are employed, with symbols 0 for event absent and 1 for event present.

This transformation converts an arithmetic process into a symbolic language that can be analyzed using information-theoretic and stochastic-process tools [5,6,7,9].

## 2.2 Twin Prime Event Language

The first event language investigated in this work is based on twin-prime occurrences. For each prime gap g(n), a binary event is defined by E_twin(n) = 1 if g(n) = 2 and E_twin(n) = 0 otherwise. For example, a sequence of prime gaps 2, 4, 2, 6, 6, 2 is represented as 1, 0, 1, 0, 0, 1. The Twin Prime Event Language is substantially denser than the Record Gap Event Language and therefore provides a useful framework for investigating local dependence, memory effects, and information propagation.

## 2.3 Record Gap Event Language

The second event language is based on record prime gaps. A record gap occurs whenever an observed prime gap exceeds all previously observed gaps. The corresponding binary language is defined by E_record(n) = 1 if g(n) establishes a new record and E_record(n) = 0 otherwise. Because record gaps become increasingly rare as larger primes are examined, the Record Gap Event Language is highly sparse. Despite this sparsity, record-gap events provide an interesting test case because they capture extreme arithmetic phenomena rather than common local events.

## 2.4 Event Density

The density of an event language is the fraction of positions containing symbol 1. For a binary sequence E(n), Density = Number of Events / Sequence Length. The event density provides a simple measure of language sparsity. For the dataset considered in this work, twin-prime events occur with a density of approximately 0.0622, while record-gap events occur with a density on the order of 10^-7. This large difference in density leads to distinct statistical characteristics.

### 2.5 Why Symbolic Languages?

The event-language representation serves two purposes. First, it transforms arithmetic phenomena into objects that can be analyzed using information theory. Quantities such as entropy, cross entropy, Markov memory, and mutual information become naturally defined [5,6,7]. Second, the representation allows arithmetic phenomena to be viewed from an informational perspective rather than solely as numerical sequences.

Numerical representations are naturally suited to studying magnitudes, distributions, and asymptotic behavior. Event-language representations are naturally suited to studying uncertainty, dependence, memory, and information propagation. Information theory is attractive in this context because it provides quantitative measures of uncertainty, dependence, and information flow that do not depend on a specific arithmetic mechanism [5,6].

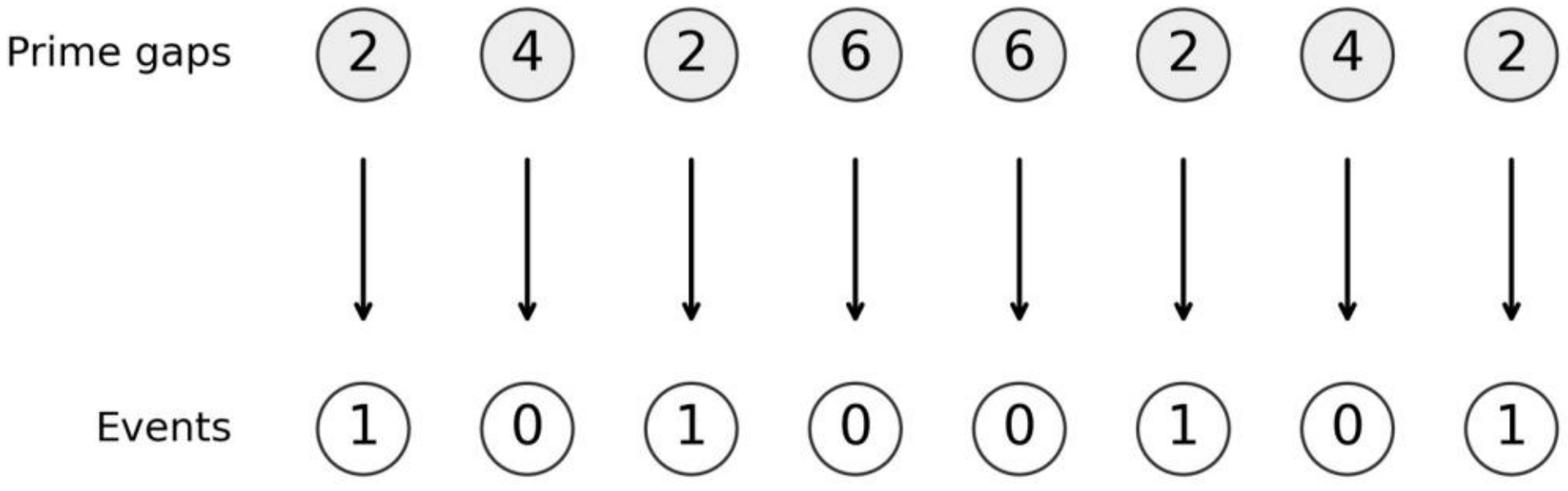

*Figure 1. Construction of the Twin Prime Event Language. Prime gaps equal to 2 are encoded as 1, while all other gaps are encoded as 0.*

## 3. Dataset and Computational Methodology

### 3.1 Prime Dataset

All analyses in this work were performed using prime numbers up to N = 5 x 10^9. The dataset contains 234,954,223 primes and therefore 234,954,222 consecutive prime gaps. The large size of the dataset allows weak statistical effects to be investigated with greater sensitivity than would be possible using smaller prime samples.

### 3.2 Event-Language Construction

Two binary event languages were constructed from the prime-gap sequence. The Twin Prime Event Language records an event whenever g(n) = 2. The Record Gap Event Language records an event whenever a newly observed prime gap exceeds all previously observed gaps.

### 3.3 Markov Modeling

Finite-order Markov models were employed to investigate memory effects [7]. For an order-k model, the probability of the next event depends on the preceding k symbols: P(E_n | E_(n-1), ..., E_(n-k)). Models of orders k = 0, 1, 2, and 3 were evaluated. Order 0 corresponds to an independent memoryless baseline.

### 3.4 Cross Entropy

Model performance was evaluated using cross entropy, H = -(1/N) sum log2 P(observed event). Lower cross entropy indicates better predictive performance [5,6]. The difference between the order-0 and higher-order models provides a quantitative measure of predictive information contained within the event language.

### 3.5 Train/Test Validation

To distinguish reproducible structure from overfitting, the event sequence was partitioned into 80% training data and 20% testing data. Markov models were constructed using only the training set and evaluated on the previously unseen test set.

### 3.6 Mutual Information

Mutual information was used to quantify statistical dependence between events separated by increasing lag distances [5,6]. For lag k, MI(k) = sum p(x,y) log2[p(x,y)/(p(x)p(y))]. Mutual information was evaluated over lags ranging from 1 to 1024.

### 3.7 Information-Horizon Analysis

To characterize the persistence of predictive information, mutual-information measurements were analyzed as a function of lag. The objective was to determine whether information decay is consistent with exponential behavior, power-law behavior, or rapid collapse toward a statistical noise floor.

### 3.8 Computational Environment

All analyses were implemented in Python and executed on commodity computing hardware. Dedicated scripts were developed for event-language construction, Markov modeling, train/test validation, mutual-information estimation, and information-horizon analysis. The workflow was designed to be reproducible and scalable to datasets containing hundreds of millions of prime gaps.

| Quantity | Value |
|---|---|
| Prime limit | 5 x 10^9 |
| Number of primes | 234,954,223 |
| Number of prime gaps | 234,954,222 |
| Twin-prime event density | 0.0622 |
| Record-gap event density | about 1.5 x 10^-7 (approximately 35 record-gap events among 234,954,222 prime gaps) |
| Training fraction | 80% |
| Testing fraction | 20% |
| Maximum MI lag analyzed | 1024 |

*Table 1. Summary of the prime dataset and event-language characteristics used throughout this study.*

## 4. Markovity Analysis

### 4.1 Objective

The objective of the Markovity analysis is to determine whether prime-derived event languages behave as memoryless processes or whether measurable dependence exists between successive events. If the event language is completely memoryless, higher-order Markov models should provide little or no predictive improvement relative to an order-0 model [7].

### 4.2 Experimental Design

Finite-order Markov models of orders k = 0, 1, 2, and 3 were constructed for the Twin Prime Event Language. For each model, cross entropy was computed over the complete event sequence.

### 4.3 Results

The Twin Prime Event Language produced the results shown in Table 2. The largest reduction in cross entropy occurs between order 0 and order 1. Orders 2 and 3 provide only negligible additional improvement

.

| Order | Cross Entropy (bits) | Information Gain |
|---|---|---|

| 0 | 0.336184 | 0 |
|---|---|---|
| 1 | 0.330224 | 0.005960 |
| 2 | 0.330216 | 0.005968 |
| 3 | 0.330215 | 0.005969 |

*Table 2. Markovity analysis results.*

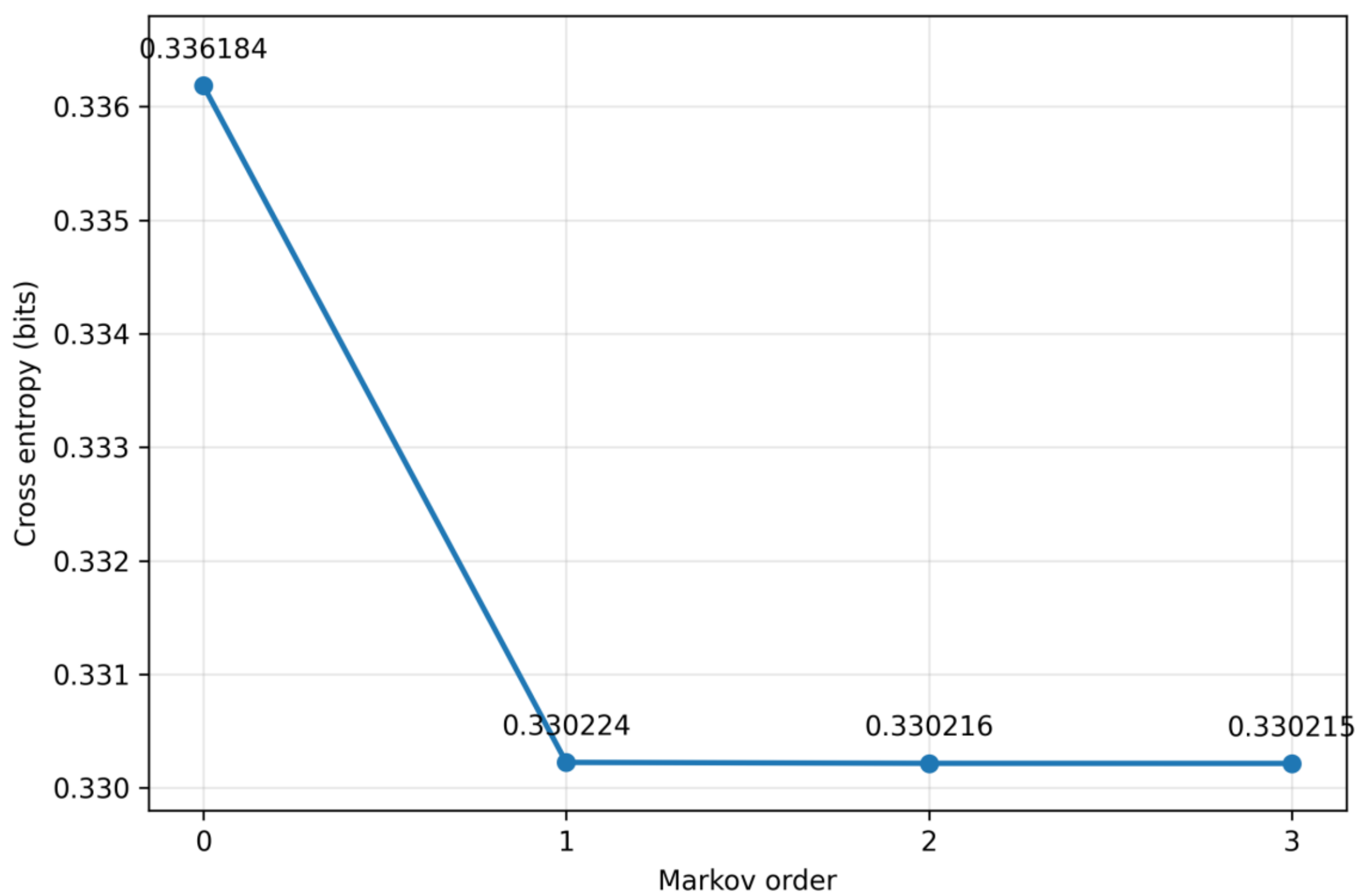

*Figure 2. Cross entropy versus Markov order. Most measurable improvement occurs between orders 0 and 1, while higher-order models provide negligible additional gain.*

### 4.4 Interpretation

The observed reduction in cross entropy indicates that the Twin Prime Event Language is not perfectly memoryless. A measurable dependence exists between neighboring events, allowing first-order models to predict future events more effectively than an independent baseline. At the same time, the very small gains achieved by orders 2 and 3 indicate that higher-order dependencies are weak. The results therefore suggest that any predictive structure present within the language is primarily local in nature.

## 5. Out-of-Sample Validation

### 5.1 Objective

The Markovity analysis demonstrated that higher-order models achieve lower cross entropy than a memoryless baseline. The train/test validation study asks whether the observed improvement reflects genuine statistical structure or merely overfitting.

## 5.2 Experimental Design

The Twin Prime Event Language was partitioned into 80% training data and 20% testing data. Markov models were constructed using only the training portion and evaluated on the held-out testing set, which contains approximately 47 million events.

## 5.3 Results

The validation results are summarized in Table 3. The order-1 model again produces a measurable reduction in cross entropy relative to the memoryless baseline, and the magnitude of the improvement is similar to that observed in the full-dataset analysis.

| Order | Test Cross Entropy (bits) | Information Gain |
|---|---|---|
| 0 | 0.325350 | 0 |
| 1 | 0.319949 | 0.005401 |
| 2 | 0.319943 | 0.005407 |

*Table 3. Out-of-sample validation results.*

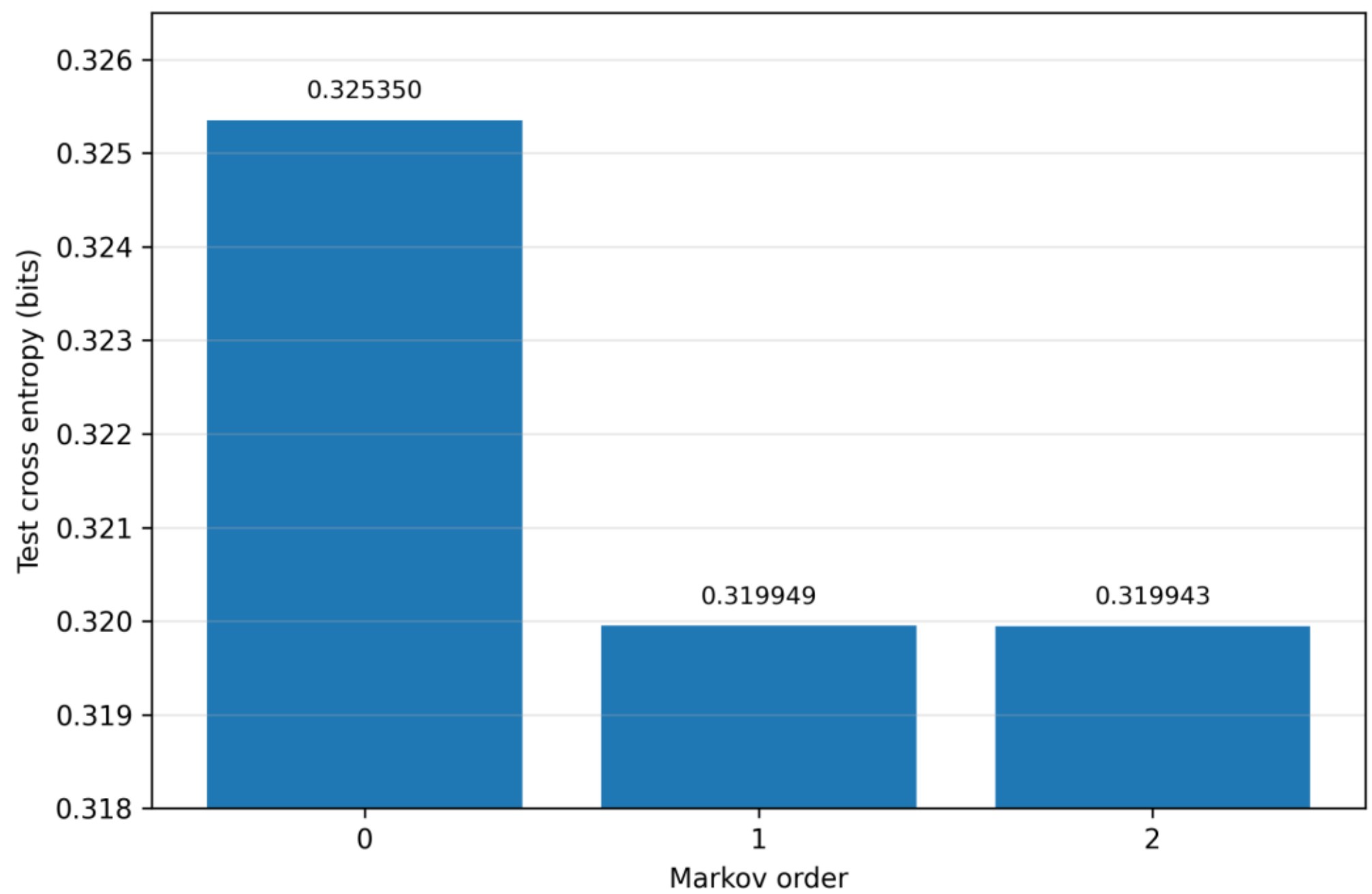

*Figure 3. Out-of-sample validation results demonstrating persistence of first-order dependence.*

## 5.4 Interpretation

The persistence of the cross-entropy reduction on the testing set strongly suggests that the observed dependence is genuine. If the improvement were caused solely by overfitting, the gain would be expected to disappear when evaluated on unseen data. The minimal improvement obtained from higher-order models indicates that nearly all predictive information is captured by first-order memory.

# 6. Mutual Information Analysis

## 6.1 Objective

The train/test validation study established that the Twin Prime Event Language contains a measurable predictive signal. The next question is how much information is contained within the language and how far that information propagates.

## 6.2 Methodology

Mutual information was computed for lags ranging from 1 to 1024. A value of zero corresponds to statistical independence, while positive values indicate measurable dependence [5,6].

## 6.3 Results

The dominant feature of the data is the dramatic difference between lag 1 and all subsequent lags. At lag 1, MI(1) is approximately 5.959 x 10^-3 bits. At lag 2, MI(2) is approximately 5.070 x 10^-7 bits. The decrease exceeds four orders of magnitude.

| Lag | MI (bits) |
|---|---|
| 1 | 5.959 x 10^-3 |
| 2 | 5.070 x 10^-7 |
| 4 | 2.123 x 10^-6 |
| 8 | 8.064 x 10^-7 |
| 16 | 4.681 x 10^-7 |
| 64 | 5.868 x 10^-8 |
| 256 | 1.888 x 10^-8 |
| 1024 | 4.938 x 10^-8 |

*Table 4. Mutual-information measurements demonstrating rapid concentration of dependence at lag 1.*

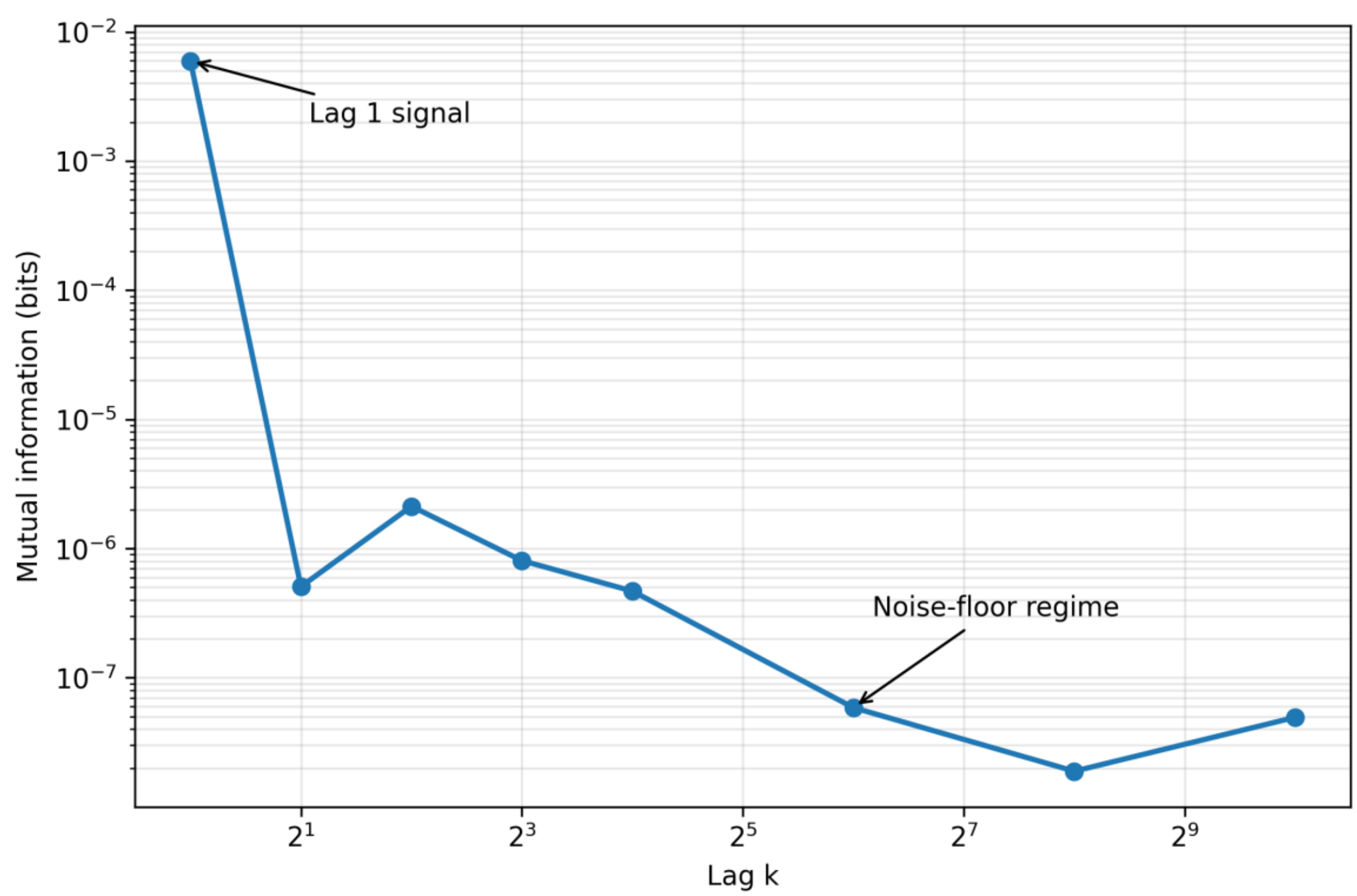

*Figure 4. Mutual information as a function of lag distance. The dominant feature is the dramatic reduction between lag 1 and lag 2, exceeding four orders of magnitude.*

### 6.4 Interpretation

The mutual-information analysis independently confirms the Markov results. The nonzero value of MI(1) demonstrates that neighboring events are not statistically independent. The abrupt reduction between lag 1 and lag 2 indicates that the overwhelming majority of measurable dependence is local. The train/test validation experiment measured a first-order information gain of approximately 0.005401 bits, while the lag-1 mutual information was approximately 0.005959 bits. Although these quantities are not identical, their similar magnitude strengthens confidence that both methods detect the same underlying statistical structure.

## 7. Information Horizon Analysis

### 7.1 Objective

The mutual-information analysis demonstrated that measurable dependence is strongly concentrated at lag 1. The information-horizon analysis asks how far predictive information persists within the Twin Prime Event Language.

### 7.2 Methodology

Mutual-information values were fitted using two decay models: MI(k) = A exp(-k/xi) and MI(k) = B k^(-alpha). Fits were evaluated using logarithmic regression and R^2.

### 7.3 Results

The fitted parameters are summarized in Table 5. The power-law model provides a somewhat better fit than the exponential model, but neither model achieves a strong coefficient of determination. The most important feature is the dramatic collapse between lag 1 and lag 2.

| Model | Parameter | Value | R^2 |
|---|---|---|---|
| Exponential | xi | 311.6 | 0.045 |
| Power law | alpha | 1.233 | 0.386 |

*Table 5. Information horizon model fits.*

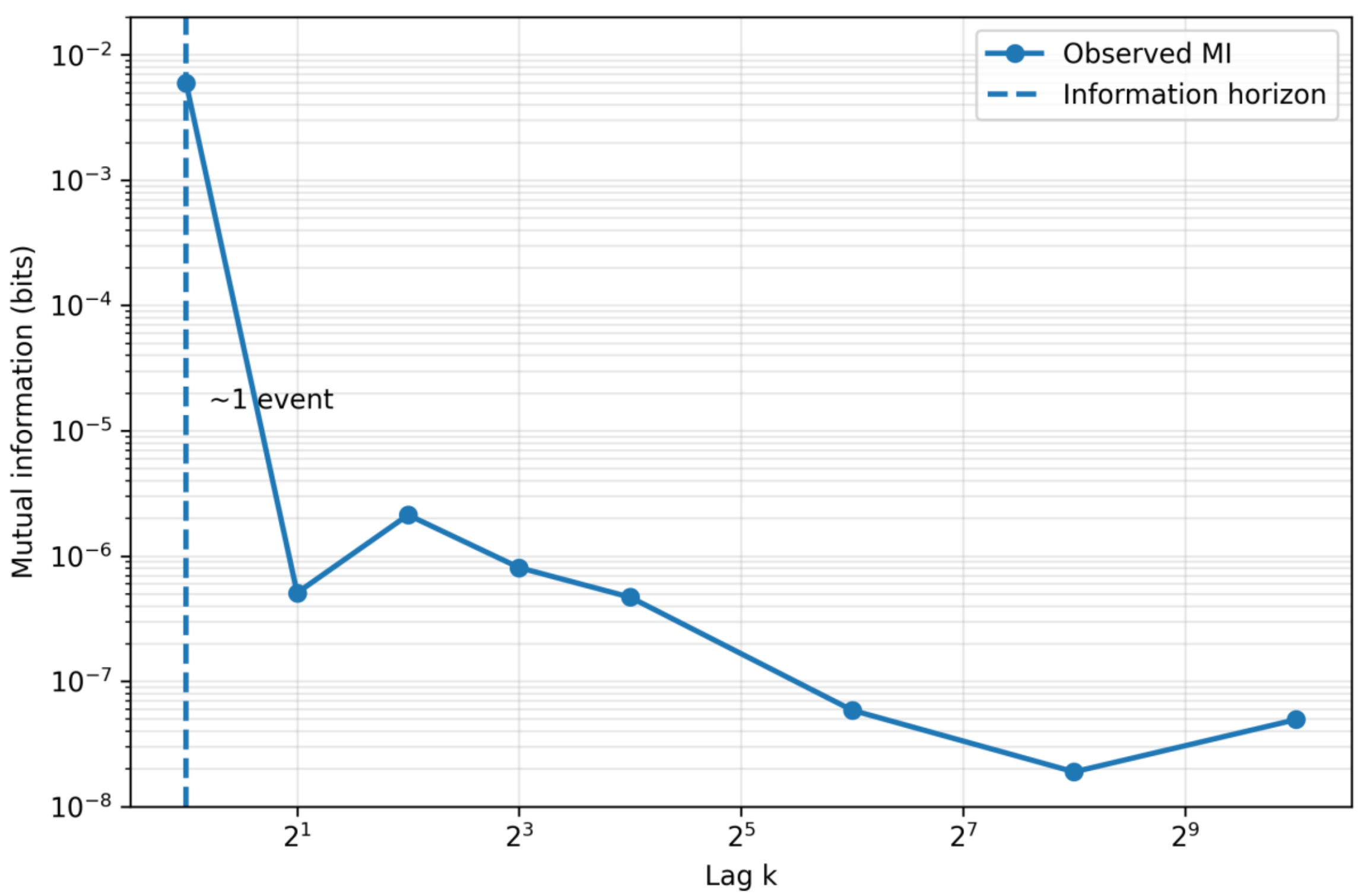

*Figure 5. Information-horizon analysis indicating that most detectable information is concentrated within a single event.*

### 7.4 Information Horizon

Neither a simple exponential decay nor a simple power-law decay provides a compelling description of the observed behavior. Instead, the data indicate a rapid collapse of information after lag 1. The information horizon is defined here as the effective distance beyond which predictive information becomes negligible relative to the dominant signal. For the Twin Prime Event Language, the information horizon appears to be approximately one event.

## 8. Discussion

### 8.1 Summary of Findings

The analyses presented in this work provide a consistent picture of the statistical structure of prime-derived event languages. The Markovity analysis demonstrated that first-order models achieve lower cross entropy than a memoryless baseline. The train/test validation study showed that this improvement survives evaluation on unseen data. Mutual-information measurements revealed that measurable dependence is concentrated almost entirely at lag 1, while the information-horizon analysis showed that predictive information decreases rapidly beyond the immediate neighborhood.

Taken together, these independent results support the conclusion that the Twin Prime Event Language possesses weak but reproducible local structure.

### 8.2 What the Results Do Not Claim

The results do not imply strong predictability of prime numbers. The observed information gains are small, and the overwhelming majority of uncertainty remains unresolved. Prime-derived event languages continue to exhibit behavior that is largely unpredictable. The present study does not propose a new theory of prime distributions. The methods employed are information-theoretic rather than number-theoretic, and the objective is descriptive rather than explanatory.

## 8.3 Representation and Hidden Structure

A central theme of this work is the role of representation. Many scientific advances arise not from new data, but from alternative representations of familiar phenomena. The event-language framework explored here follows the same philosophy. The prime numbers themselves are unchanged. The prime gaps themselves are unchanged. What changes is the perspective through which these phenomena are examined. By representing arithmetic events as symbolic languages, concepts such as entropy, Markov memory, mutual information, and information propagation become naturally defined [1,2,3,7].

### 8.3.1 Prime Numbers as Events

Traditional number theory studies prime numbers primarily as numerical objects characterized by magnitude, density, distribution, and asymptotic behavior [1-4]. The Prime Event Language framework introduces a complementary perspective in which arithmetic phenomena are viewed as event-generating processes. This representational shift does not alter the underlying mathematics. The prime numbers remain unchanged, and the prime-gap sequence remains unchanged. What changes is the set of quantities that become naturally observable and measurable.

When arithmetic phenomena are represented as symbolic event sequences, concepts such as entropy, mutual information, predictive information, information horizons, and Markov structure become directly accessible [5,6,7,9]. From this perspective, arithmetic systems may be studied not only as collections of numbers but also as generators of information-bearing event sequences. The principal contribution of the Prime Event Language framework is therefore not the introduction of a new mathematical theorem, but the introduction of a new representational framework capable of exposing measurable information-theoretic structure within familiar arithmetic phenomena.

## 8.4 Information-Theoretic Perspective

From an information-theoretic perspective, the most significant finding is not merely that dependence exists, but that the dependence is highly localized. The close numerical agreement between the first-order Markov information gain (0.005401 bits) and the lag-1 mutual information (0.005959 bits) suggests that both analyses are detecting the same underlying local statistical dependence. The resulting picture is one of weak local organization rather than substantial long-range memory.

## 8.5 Limitations

Several limitations should be acknowledged. First, the present study focuses primarily on the Twin Prime Event Language. Although the Record Gap Event Language is introduced as part of the framework, its extreme sparsity limits the range of statistical analyses that can currently be performed. Second, the observed information gains are small in absolute magnitude. While statistically reproducible, they do not imply substantial predictability of prime-derived events. Third, the analyses presented here are descriptive rather than explanatory. They identify measurable information structure but do not provide a number-theoretic mechanism responsible for the observed dependence.

Although the detected dependence is statistically reproducible, its magnitude remains small and should not be interpreted as evidence of strong predictability or as a challenge to existing probabilistic models of prime distributions [4,9]. In addition, the present analysis is empirical and does not establish a theoretical mechanism responsible for the observed dependence.

## 8.6 Future Directions

Additional arithmetic event languages could be constructed from larger prime constellations, modular residue patterns, residue-class dynamics, or composite arithmetic events. Future work may also explore entropy-rate estimation, higher-order event languages, compression-based complexity measures, and connections to algorithmic information theory [10].

Event languages may also be combined with network-theoretic representations to investigate whether information flow and topological organization are related. In particular, event-transition graphs could be analyzed for clustering, average path length, modularity, and small-world structure. Such extensions would connect Prime Event Languages to broader questions about topology, information processing, and complex systems.

## 8.7 Discussion Summary

The Twin Prime Event Language is neither perfectly memoryless nor strongly predictable. Instead, it exhibits weak but reproducible local structure that becomes visible when arithmetic events are represented as symbolic languages and analyzed using information-theoretic tools. The broader significance of the present work lies not in the magnitude of the observed dependence, but in the perspective through which it becomes visible.

| Scientific Question | Method | Result |
| --- | --- | --- |
| Is the language memoryless? | Markovity Analysis | No |
| Does dependence survive validation? | Out-of-Sample Validation | Yes |
| Where is information concentrated? | Mutual Information Analysis | Lag 1 |
| Information horizon? | Information Horizon Analysis | Approximately 1 event |
| Long-range memory present? | Combined Analysis | Extremely weak |
| Why is structure detectable at all? | Event-Language Representation | Hidden information-theoretic structure becomes measurable |

*Table 6. Integrated findings.*

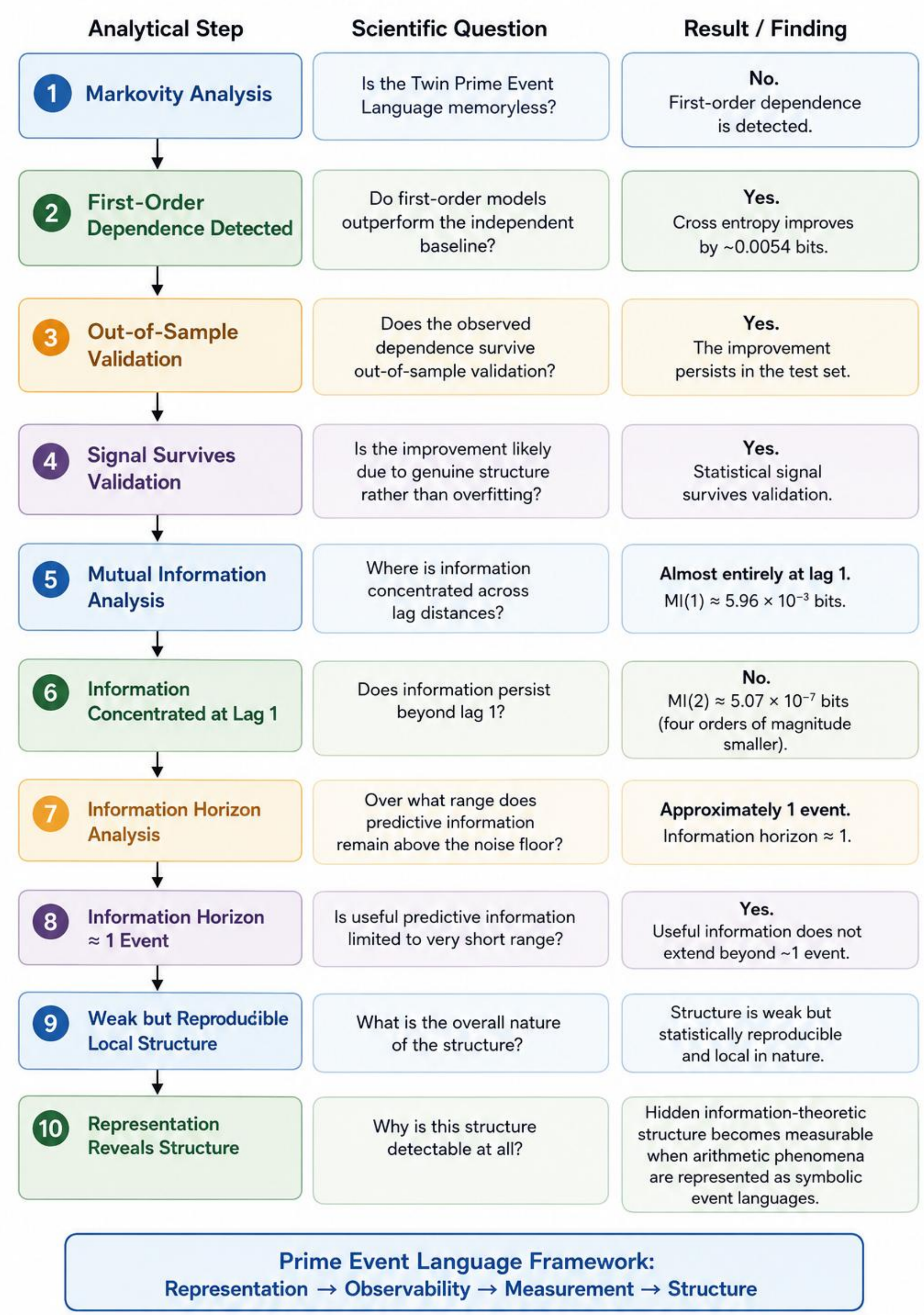

*Figure 6. Summary of analytical steps and scientific findings. The Prime Event Language framework reveals weak but statistically reproducible local structure, concentrated at lag 1, with an information horizon of approximetely one event. Most importantly, the representation itself makes hidden structure measurable.*

.

### 8.8 Broader Implications

The Prime Event Language framework is motivated by a general scientific principle: representation influences observability. Different representations of the same underlying phenomenon may reveal different classes of measurable structure.

In the present study, transforming prime-gap data into symbolic event sequences makes quantities such as entropy, mutual information, predictive information, and information horizons naturally observable [5,6,9]. Similar representational transformations have historically played important roles in physics, information theory, biology, and computer science, where alternative descriptions often reveal organizational principles that are difficult to identify within the original formulation.

The broader significance of the present work therefore extends beyond twin-prime events themselves. The central question is whether arithmetic systems possess informational organization that becomes measurable only after an appropriate representational transformation. The results presented here provide preliminary evidence that such an approach may be fruitful and motivate further exploration of information-theoretic representations of mathematical objects.

### 8.9 Reproducibility Statement

All analyses reported in this work were implemented in Python 3.14 using custom-developed scripts for prime-gap extraction, event-language construction, Markov modeling, train/test validation, mutual-information estimation, and information-horizon analysis. The computational workflow was designed to be fully reproducible and scalable to datasets containing hundreds of millions of prime gaps.

Source code, configuration files, generated datasets, and figure-generation scripts will be released through a public repository to facilitate independent verification and future extensions of the Prime Event Language framework. All numerical results reported in this study were generated from deterministic computations applied to the complete prime dataset up to N = 5 x 10^9.

### 8.10 Statistical Robustness

Although the observed information gains are small in absolute magnitude, the dataset analyzed in this study contains 234,954,222 prime gaps, providing substantial statistical power. The persistence of the signal under train/test validation strongly suggests that the observed dependence is not an artifact of overfitting. Furthermore, the close numerical agreement between the first-order Markov information gain and the lag-1 mutual-information measurement provides convergent evidence from two independent analytical methodologies.

Taken together, these observations indicate that the detected dependence is statistically robust despite its weak magnitude. The results therefore support the conclusion that the Twin Prime Event Language exhibits genuine but highly localized informational structure.

## 9. Conclusions

This work introduced the concept of a Prime Event Language, in which arithmetic phenomena are represented as symbolic event sequences and analyzed using tools from information theory and stochastic processes. Using all primes up to N = 5 x 10^9, two event-language representations were constructed from prime-gap data: the Twin Prime Event Language and the Record Gap Event Language. The primary objective was to investigate whether measurable information structure becomes visible when prime-derived phenomena are viewed through an event-language framework.

The Markovity study demonstrated that first-order models outperform a memoryless baseline. Train/test validation confirmed that this predictive advantage survives evaluation on unseen data. Mutual-information analysis revealed that measurable dependence is concentrated almost entirely at lag 1. Information-horizon analysis showed that predictive information decreases by more than four orders of magnitude after a single lag step and becomes negligible beyond the immediate neighborhood.

The principal findings are: (1) prime-derived event languages are not perfectly memoryless processes; (2) measurable first-order dependence exists within the Twin Prime Event Language; (3) the dependence survives out-of-sample validation; (4) most predictive information is concentrated in the immediately preceding event; and (5) the effective information horizon is approximately one event.

The contribution of this work is not the introduction of new mathematical methods. Rather, it is the application of established information-theoretic tools to a new representation of prime-derived arithmetic phenomena. In summary, prime-derived event languages appear to occupy an intermediate regime between complete independence and substantial predictability. Their statistical structure is weak, highly localized, and detectable only through large-scale analysis. Nevertheless, the results demonstrate that meaningful information-theoretic organization becomes measurable when arithmetic phenomena are viewed through an event-language representation.

Scientific progress often arises not only from new data or new methods, but also from new representations that make previously hidden structure measurable. Future work may explore whether similar event-language representations reveal information structure in other arithmetic systems, including prime constellations, residue-class dynamics, and composite arithmetic events.